\documentclass[runningheads]{rnl}
\usepackage[latin1]{inputenc}
\usepackage[english]{babel}
\usepackage{graphicx,epsf,subfigure}  

\begin{document}
 
\title*{Etude d'une équation de convection-réaction-diffusion en écoulement compressible}
 
\titretab{début du titre
\protect\newline fin du titre}

\titrecourt{Etude d'une équation ARD compressible}
 
\author{F. Bianco\inst{1}
\and S. Chibbaro\inst{2}
\and R. Prud'homme\inst{3}} 
\index{Nom prénom}              

\auteurcourt{F. Bianco {\it et al}.}
 
\adresse{Institut Jean le Rond d'Alembert
Université Pierre et Marie Curie,
4 place Jussieu, 75252 Paris Cedex 05
}

\email{bianco@dalembert.upmc.fr}

\maketitle


\begin{resume}
Nous avons \'etudi\'e la propagation d'un front de flamme dans une configuration unidimensionnelle en solvant num\'eriquement une \'equation de advection-diffusion-r\'eaction (ADR). Nous avons choisi un mod\`ele simplifi\'e dans lequel le couplage avec d'autres phenom\`enes est n\'eglig\'e et le fluide r\'eactant est une mixture de gaz. En outre, la vitesse de l'\'ecoulement compressible est donn\'ee de façon analytique. 
A l'aide de la simulation num\'erique nous analysons la différence entre des modéles qui sont communement utilis\'es dans les \'etudes fondamentales en combustion et en biologie. Puis nous avons caracteris\'e l'effet de la compressibilit\'e sur la dynamique du front et dans quelles conditions il est possible que la flamme s'eteigne avant la saturation de la r\'eaction.
\end{resume}

\begin{resumanglais}
We have studied the front propagation in a one dimensional case of combustion by solving numerically 
 an advection-reaction-diffusion equation. The physical model is siplified so that no coupling phenomena are considered and 
the reacting fluid is a binary mixture of gases. The compressible flow field is given analitycally. 
We analyse the differences between popular models used in fundamental studies of compressible combustion and biological problems. Then, we investigate the effects of compressibility on the front interface dynamics for different reaction types and we characterise the conditioins for which the reaction stops before its completion.
\end{resumanglais}

\section{Introduction}
Transport of reacting species advected by laminar or turbulent flows described by an advection reaction diffusion equation (ARD),
is an issue of interest in many fields, e.g., population dynamics, propagation of plankton in oceanic currents,
 reacting chemicals in the atmosphere, ozone dynamics, complex chemical reactions, and combustion.
While these phenomena have been and are being widely studied in the case of incompressible flows, 
the transport of reactive species advected by compressible flows is less discussed. 
Only recently has been extended to the study of population dynamics \cite{Bianco_Federico1}.
Combustion processes are particularly complex because they involve a large number of chemical species
and a large number of reactions. In addition, the concentration of these species, and the release of chemical energy during
reactions, greatly affect the flow field of the fuel and oxidizer mixture. Consequently, the numerical study of this phenomenon
requires the solution of a large number of coupled partial difference equations (conservation of species, momentum conservation
  and energy conservation).\\
The complexity can be greatly reduced if we minimize the physical details. In particular, if we neglect the effects
  of the coupling between the various equations and we minimize the number of species, combustion can be
described approximately by an advection reaction diffusion equation:
\begin{equation}
 \frac{\partial{ \rho \phi}}{\partial t}+{\bf \nabla} \cdot (\rho {\bf u} \phi)= \nabla \cdot (\gamma \nabla \phi)+ s
\label{Bianco_Federcoeq1}
\end{equation}
which describes the spatio-temporal behavior of a fraction $\phi$ of the reactive mixture which moves with velocity ${\bf u}$.
The mean density of the mixture is $\rho$, $\gamma=\rho D$ is the diffusion coefficient and $s$ is the source term.  
\\ Although there is a strong connection between the study of population dynamics
  and combustion, as the simplified mathematical model is equivalent when we consider incompressible flows, in
 the case of compressible flows we will show briefly that the parallelism requires a little more attention.

\section{Model}
If we consider a compressible reactive flow, and if the  
fluid is a mixture of $N$ perfect gases, the local total mass balance over a volume $V$ gives:
\begin{equation}
\frac{\partial{\rho}}{\partial t} + {\bf \nabla} \cdot (\rho{\bf u}) = 0
\label{Binco_Federicoeq2}
\end{equation}  
where $\rho({\bf x},t)=\sum_k \rho_k $ is the density of the mixture and ${\bf u}({\bf x},t)$ is the
 resultant velocity field.\cite{Bianco_Federico2}\cite{Bianco_Federico3}
This equation is the summation over the $N$ mass species conservation:
\begin{equation}
\frac{\partial{\rho c_k}}{\partial t} + {\bf \nabla} \cdot \left[ \rho ({\bf u}+{\bf v}_k) c_k\right]= R_k~~,~~k=1,...,N
\label{Bianco_Federicoeq3}
\end{equation}
where $c_k({\bf x},t)$ is the mass fraction ($\rho_k/\rho$), ${\bf v}_k({\bf x} ,t)$ and $R_k$ are the diffusion velocity
 and the reaction rate of the $k^{th}$ species respectively.
Of course, by definition, $\sum_{k=1}^N R_k=\sum_{k=1}^N {\bf v}_k c_k=0$ as mass cannot be generated during chemical reactions.\\
If the mixture contains only two species, the pressure gradients are small, and volume forces are neglected \cite{Bianco_Federico2}
equation (\ref{Bianco_Federicoeq3}) for the $k^{th}$ species can be expressed as:
\begin{equation}
\rho \frac{d{ c_k}}{d t}=\rho \left[\frac{\partial{ c_k}}{\partial t} + {\bf u} \cdot {\bf \nabla} 
 c_k  \right]= \nabla \cdot (\rho D \nabla c_k)+ R_k~~,~~k=1,2.
\label{Bianco_Federicoeq4}
\end{equation}
where the diffusion velocity is rewritten according to Fick's law.\\
Equations (\ref{Bianco_Federicoeq4}) describe the spatial, and time, behavior of the fractions ($c_1$,$c_2$)
 of a binary reacting mixture in a compressible velocity field.
Such as, for example, the fresh air as well as the burnt gases in a combustion.\\
If $D$ is constant, equation (\ref{Bianco_Federicoeq4}) in conservative form becomes: 
\begin{equation}
 \frac{\partial{\rho c_k}}{\partial t} +  {\bf \nabla} \cdot (
\rho {\bf u} c_k)  =D \nabla \cdot(\rho \nabla c_k)+ R_k~~,~~k=1,2.
\label{Bianco_Federicoeq5}
\end{equation}
If $D$ and $\rho$ are constant, equation (\ref{Bianco_Federicoeq4}) become the widely studied incompressible advection-reaction-diffusion
equation with constant diffusion coefficient.\\
Equations (\ref{Bianco_Federicoeq4}) and (\ref{Bianco_Federicoeq5}) differ from the equation:
\begin{equation}
 \frac{\partial{\theta}}{\partial t} +  {\bf \nabla} \cdot ({\bf u} \theta)  =D_0 \nabla^2 \theta+ F(\theta)
\label{Bianco_Federicoeq6}
\end{equation}
which is often find in literature \cite{Bianco_Federico1}\cite{Bianco_Federico4} to describe the advection, diffusion and reaction of a scalar 
in a compressible flow. This equation is typical in the study of population dynamics and 
the scalar $\theta({\bf x},t)$, is the concentration
of a population \cite{Bianco_Federico1}. 
However, this model is not correct for the concentration
of the combustion products. In fact, in equation (\ref{Bianco_Federicoeq6}), if ${\bf \nabla }\cdot{\bf u}\ne 0$ the concentration $\theta$ can take 
values greater then one because is not a fractional parameter. Notably if one considers the Fisher-Kolmogorov-Petrovskii-Piskunov reaction rate (FKPP)\cite{Bianco_Federico5}, the rate of grow ($F(\theta)=\alpha c(1-c)$) in equation (\ref{Bianco_Federicoeq6}), if not correctly
rescaled, can take negative values. This could be in agree with the population dynamics point of view but is formally 
incorrect for the chemical reactions during a combustion.

\subsection{Problem statement and numerical details}
\begin{figure}[htbp]                    
  \begin{center}
    \begin{tabular}{ccc} 
      \setlength{\epsfysize}{4.2cm}
      \subfigure[]{\epsfbox{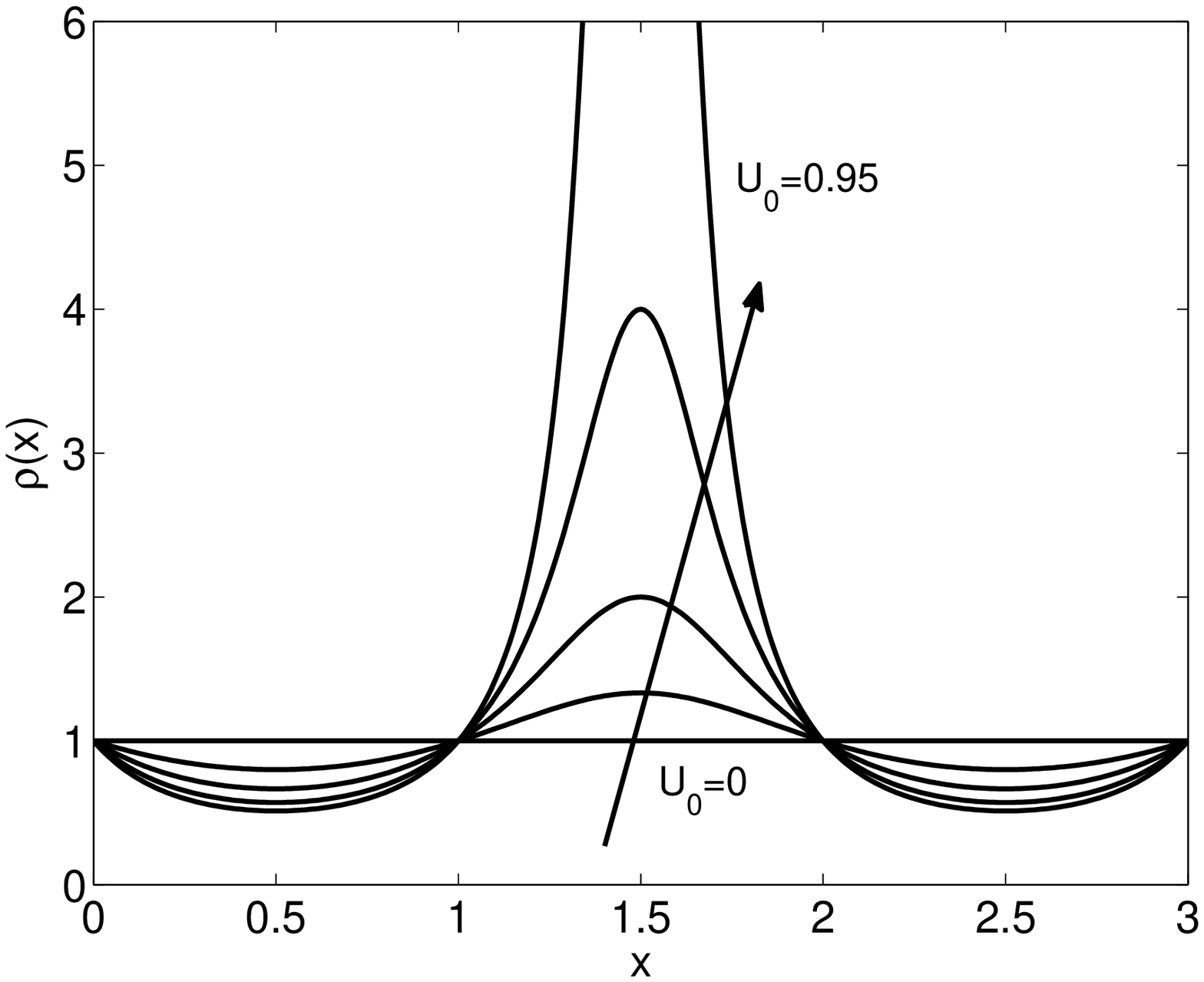}} &
      \setlength{\epsfysize}{4.2cm}
      \subfigure[]{\epsfbox{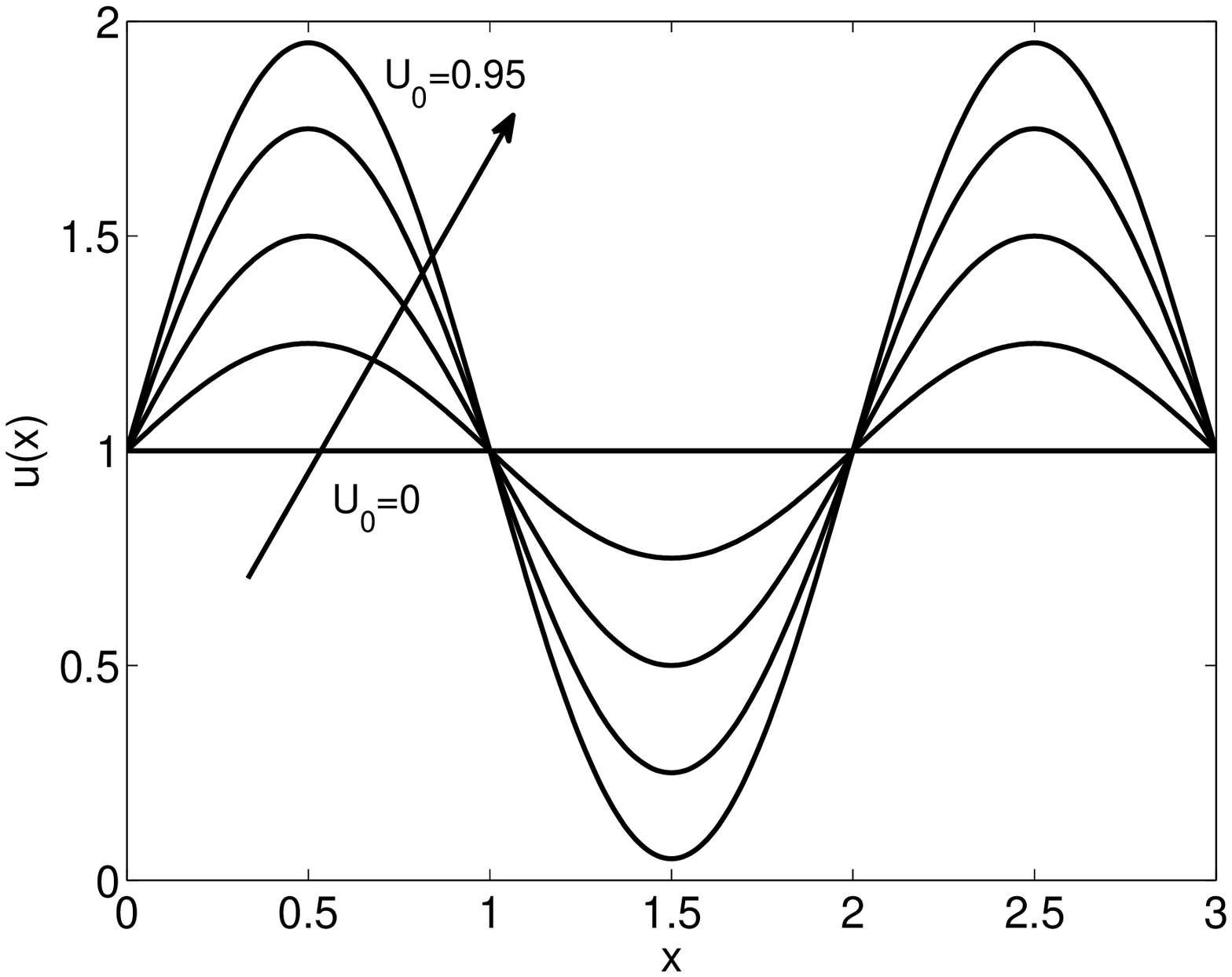}} &
      \setlength{\epsfysize}{3.95cm}
      \subfigure[]{\epsfbox{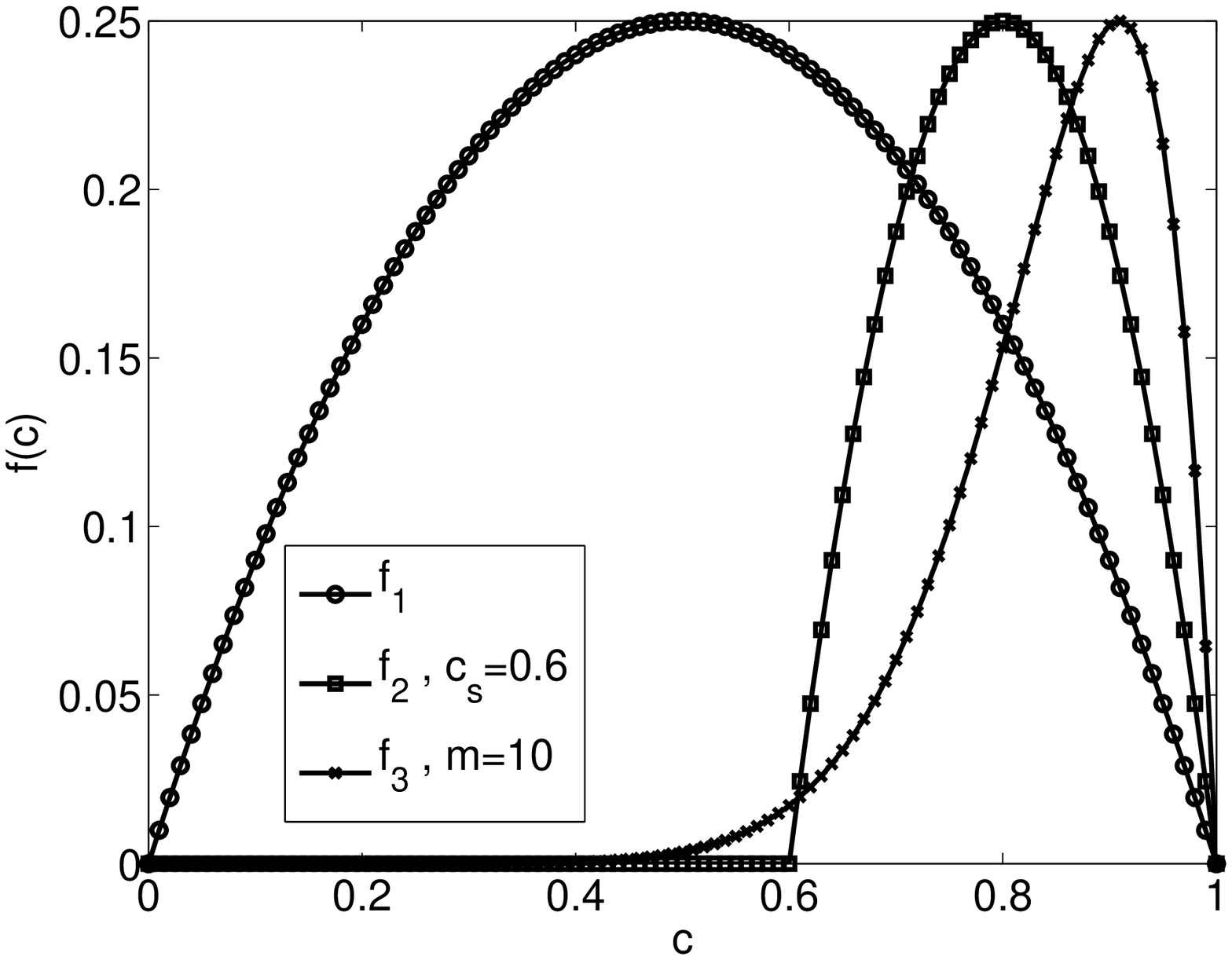}} \\
    \end{tabular}
    \caption{Density (a) and velocity field (b) of the mixture as function of the parameter $U_0$. 
In figure (c) the three rate of reaction considered, as function of the mass fraction: the FKPP 
reaction ($f_1$); the ignition type reaction ($f_2$) with threshold $c_s=0.6$ and the general mth order 
Fisher reaction ($f_3$) with $m=10$.  }
    \label{Bianco_Federicofig1}             
  \end{center} 
\end{figure}

We have considered for simplicity a one dimensional case in which there is no feedback from the reaction rate,
to the velocity field of the mixture. Moreover, the mixture is sufficiently dilute to neglect the heat release and the
 variation of temperature, thus we do not need to consider separately the energy equation. 
The velocity field is steady state and analytical so that the total mass balance becomes:
\begin{equation}
 \frac{\partial{\rho u}}{\partial x} =0~~\rightarrow~~\rho u=\rho \left[ 1+U_0 \sin(\frac{\pi x}{L}) \right]=constant=\beta=1
\label{Bianco_Federicoeq7}
\end{equation}
where $U_0$ is a parameter defined on the interval $[0,1]$ and $L=1$ is the length scale of the velocity field. As shown in Fig.\ref{Bianco_Federicofig1},
when $U_0=0$ the mixture moves with constant velocity, the density is constant and the problem becomes incompressible. On the other
hand if $U_0$ tends to one the density varies very rapidly between $0.5$ and infinity.
\\If we consider the product $\rho D=\gamma$ constant and we collect and replace the velocity $u$ with its analytic function, equation (\ref{Bianco_Federicoeq4}) becomes:
\begin{equation}
 \frac{\partial{c}}{\partial t}     = \left[ 1+U_0 \sin(\frac{\pi x}{L})\right]\left(\frac{\gamma}{\beta}  \frac{\partial^2{c}}{\partial x^2}-\frac{\partial{c}}{\partial x} \right) +  f(c)
\label{Bianco_Federicoeq8}
\end{equation}
In this case, and later on, we will refer to $c(x,t)$ as the mass fraction of the reacted products, like the mass fraction of burnt gases in a combustion reaction. 
\\If only $D$ is constant the balance equation for the burnt gases (\ref{Bianco_Federicoeq5}) gives:
\begin{equation}
 \frac{\partial{c^+}}{\partial t}    =u \left[\frac{D}{\beta} \frac{\partial}{\partial x} \left(\rho \frac{\partial c^+ }{\partial x}\right) -\frac{\partial{c^+}}{\partial x} \right]+  f(c^+)
\label{Bianco_Federicoeq9}
\end{equation}
If we use equation (\ref{Bianco_Federicoeq6}) instead, we will get the following one dimensional ARD equation:
\begin{equation}
 \frac{\partial{\theta}}{\partial t}    = D \frac{\partial^2 \theta }{\partial x^2}  -u \frac{\partial{\theta }}{\partial x} -\theta \frac{\partial{u }}{\partial x}  +  f(\theta)
\label{Bianco_Federicoeq10}
\end{equation}
\\We will consider three different non-linear
 rate of reaction, as shown by Fig.\ref{Bianco_Federicofig1}.c\\
The FKPP non linear rate of reaction or autocatalytic reaction:
\begin{equation}
f_1(c)=\alpha_1 c(1-c)
\end{equation}
The ignition type rate of reaction:
\begin{equation}
\displaystyle
  f_2(c)=\left\{ \begin{array}{cc}
		    0 			 & if~~c\le c_s\\
		    \alpha_2(c-c_s)(1-c) & if~~c>c_s 
		  \end{array} \right.
\end{equation}
And the general $m^{th}$ order Fisher's non-linearity:
\begin{equation}
f_3(c)=\alpha_3 c^m(1-c)
\end{equation}
If we use equations (\ref{Bianco_Federicoeq8}) and (\ref{Bianco_Federicoeq9})  $c$ is defined in the interval $[0,1]$. 
The reaction rates then, by definition, are always positives or zero which means that the reaction is irreversible.  
As we will show soon this is not true for equation (\ref{Bianco_Federicoeq6}).\\
The constants $\alpha_1$, $\alpha_2$ and $\alpha_3$ are chosen so that $max\left(f_i(c)\right)=\alpha_1/4$. 
In this way reactions have  different reaction rates but with
comparable characteristic times.\\
Equation (\ref{Bianco_Federicoeq8}), (\ref{Bianco_Federicoeq9}) and (\ref{Bianco_Federicoeq10}) have been 
solved numerically using a fourth order finite different scheme for spatial discretization
and a fourth order Runge-Kutta method for time advancement. The numerical domain has a dimension $L_x=nL$,
 where $n$ is an even natural number,  and periodic boundary
condition have been applied.
\section{Results}
\begin{figure}[htbp]                    
  \begin{center}
    \begin{tabular}{ccc}
      \setlength{\epsfysize}{4.2cm}
      \subfigure[]{\epsfbox{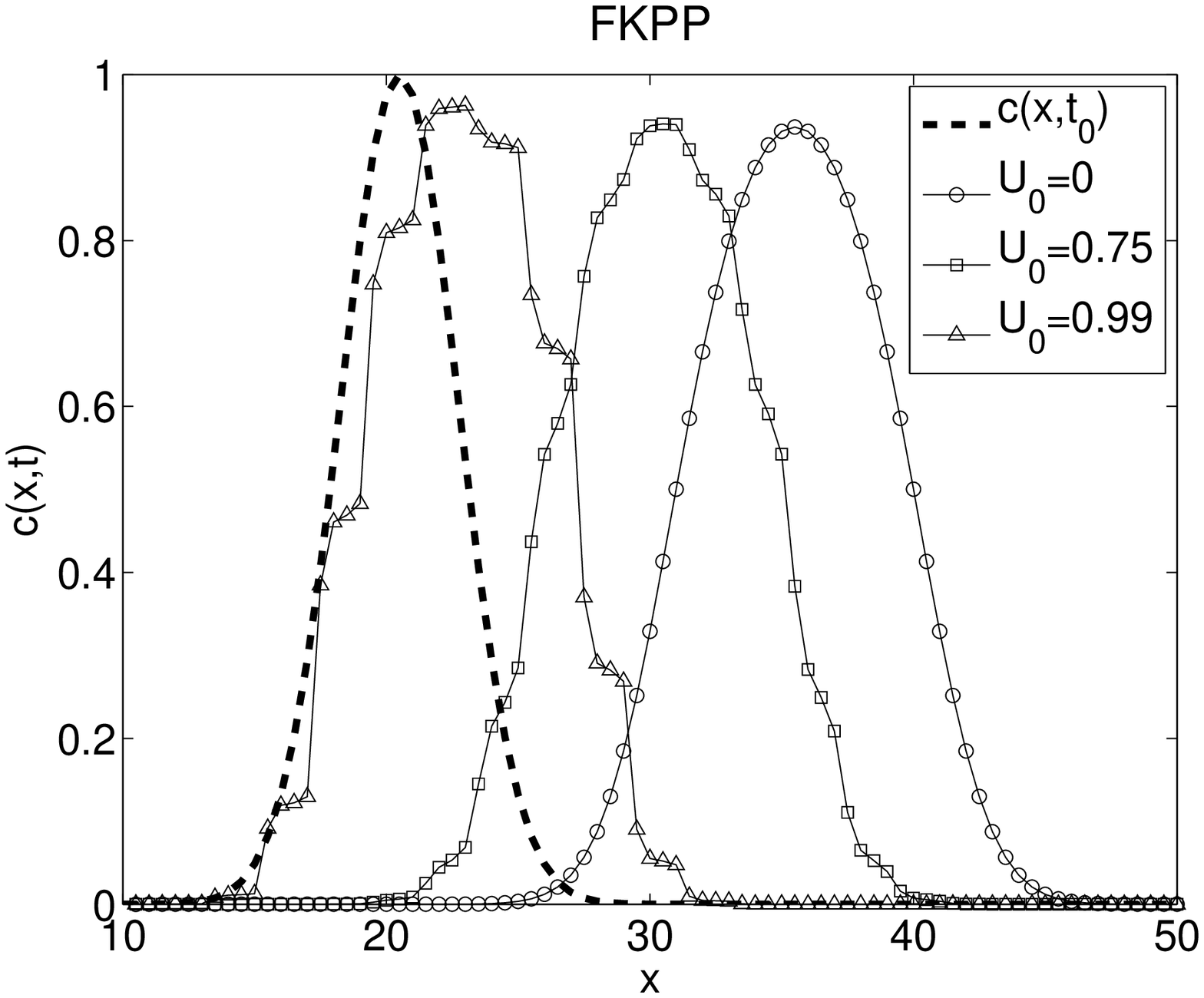}} &
      \setlength{\epsfysize}{4.2cm}
      \subfigure[]{\epsfbox{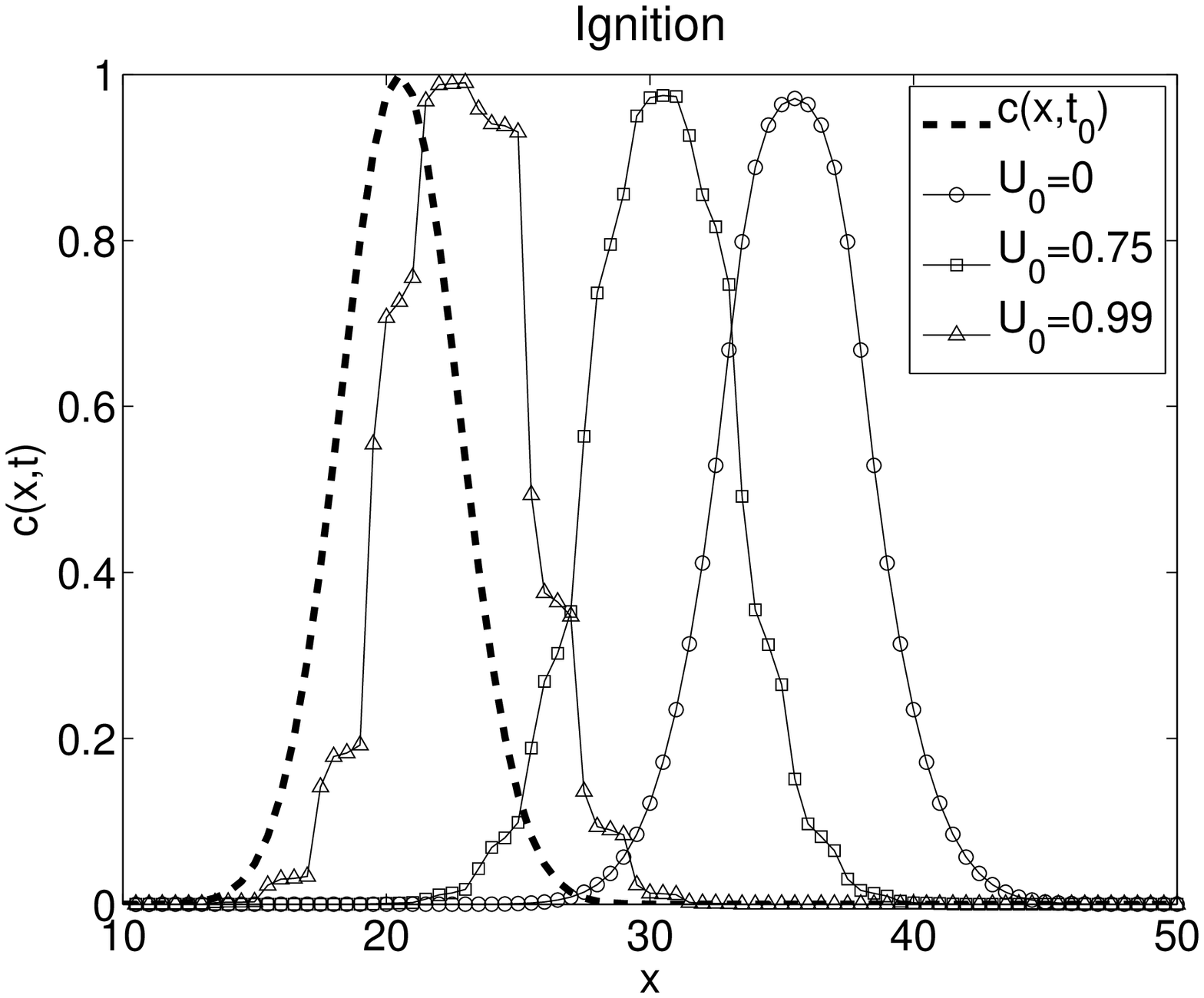}} &
      \setlength{\epsfysize}{4.2cm}
      \subfigure[]{\epsfbox{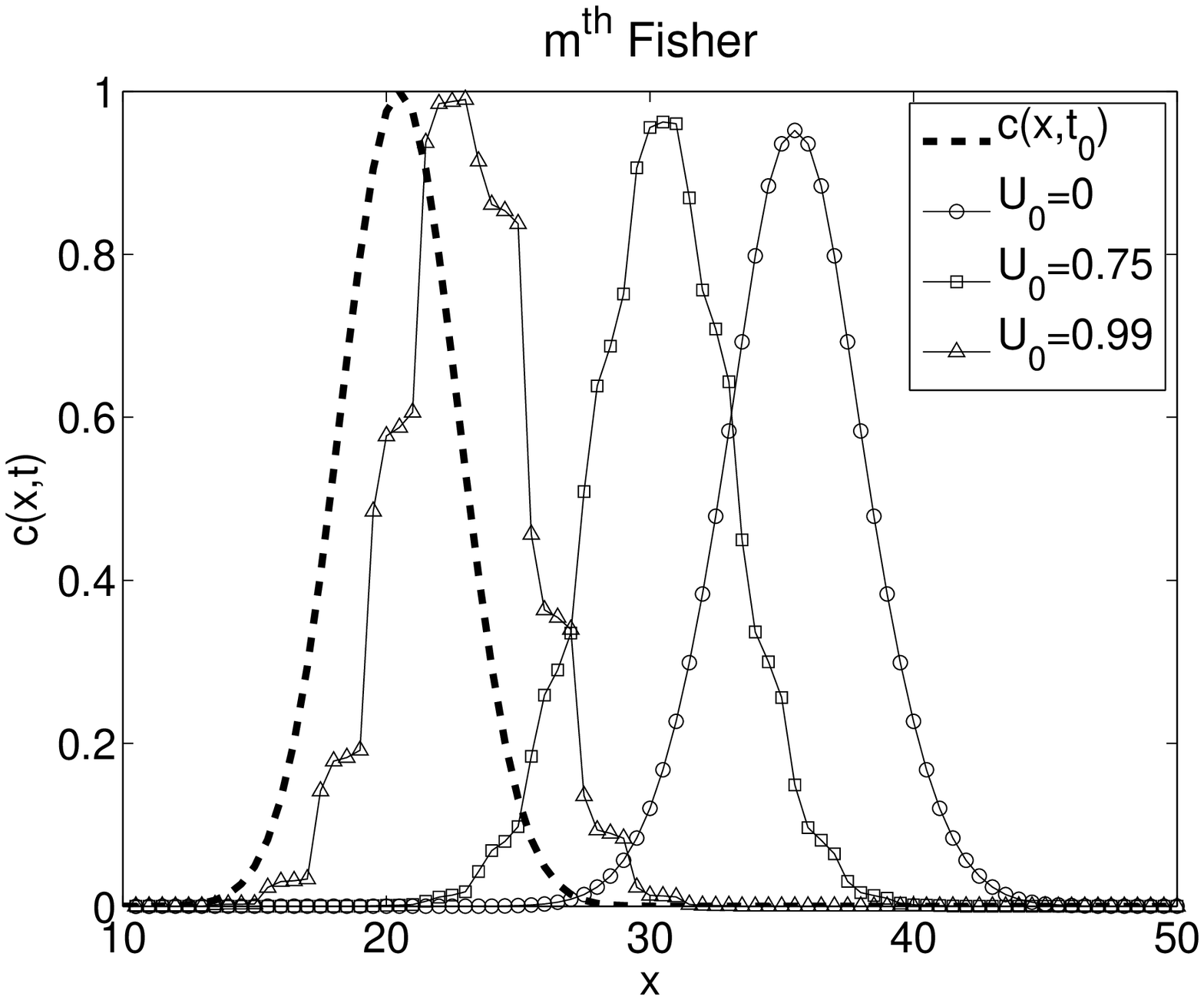}} \\[-0.2cm]
    \end{tabular}
    \caption{In contiunous line, the mass fraction of burnt gases after a transient of $t=15$ in the case of FKPP (a), ignition (b) 
and generalized mth order Fisher reaction (c). In dashed line the initial condition where $x_0=20.5$ and $\sigma_0=10$. 
The parameter $\frac{\gamma}{\beta}=0.1$ ($\beta=1$,$\gamma=0.1$) and the coefficient $\alpha_1=0.1$ are such that the maximum
 rate of reaction is $max(f_i)=0.025$. In the case of ignition reaction the threshold was set up $c_s=0.6$ while the exponent on the
general Fisher non linearity is $m=10$.
 Having fixed the grid step $dx=0.025$ and the time step $dt=0.0005$ we have run different simulations  varying
 the coefficient $U_0$. }
    \label{Bianco_Federicofig2}             
  \end{center}
\end{figure}

\begin{figure}[htbp]                    
  \begin{center}
    \begin{tabular}{ccc} 
      \setlength{\epsfysize}{4.2cm}
      \subfigure[]{\epsfbox{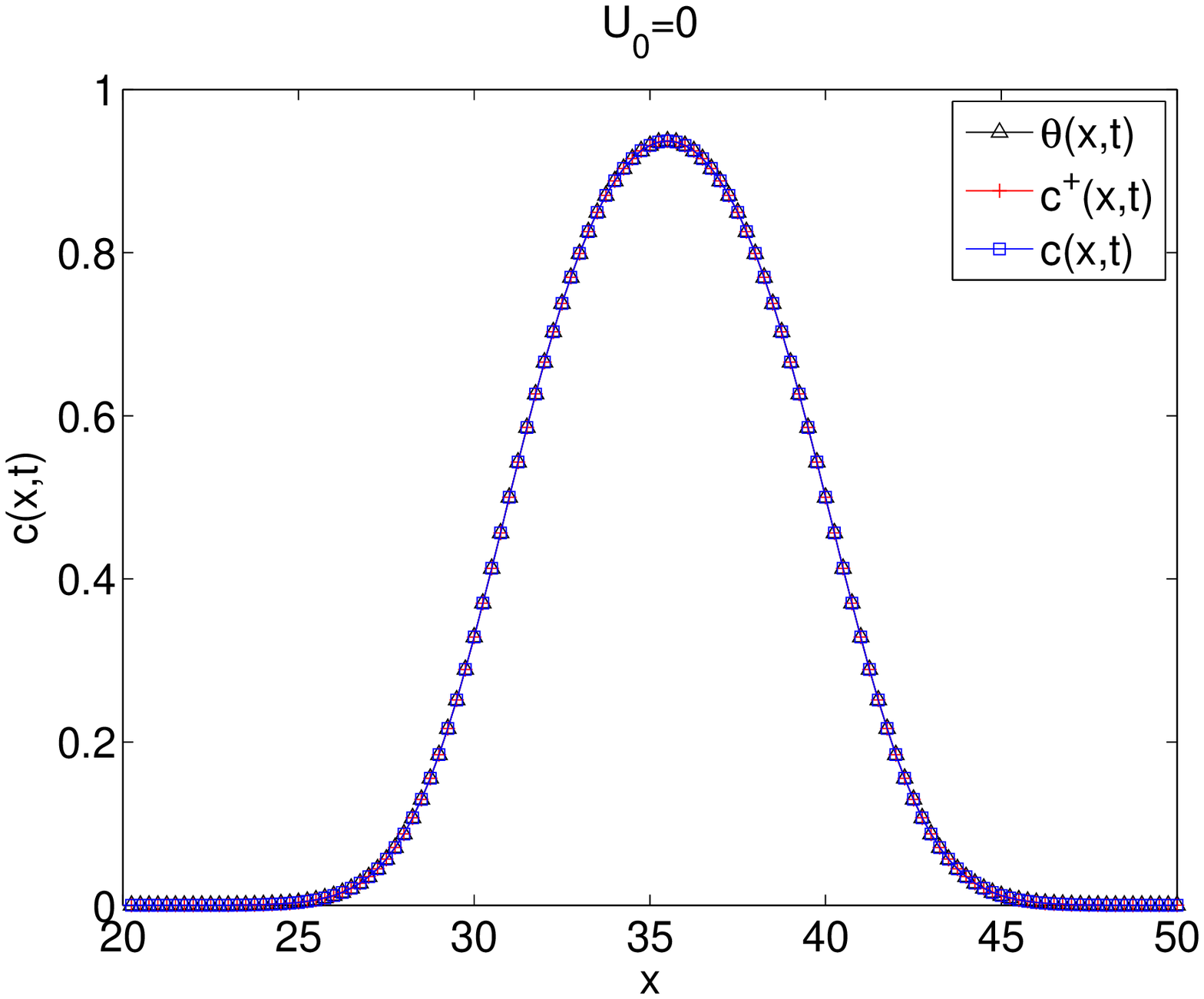}} &
      \setlength{\epsfysize}{4.2cm}
      \subfigure[]{\epsfbox{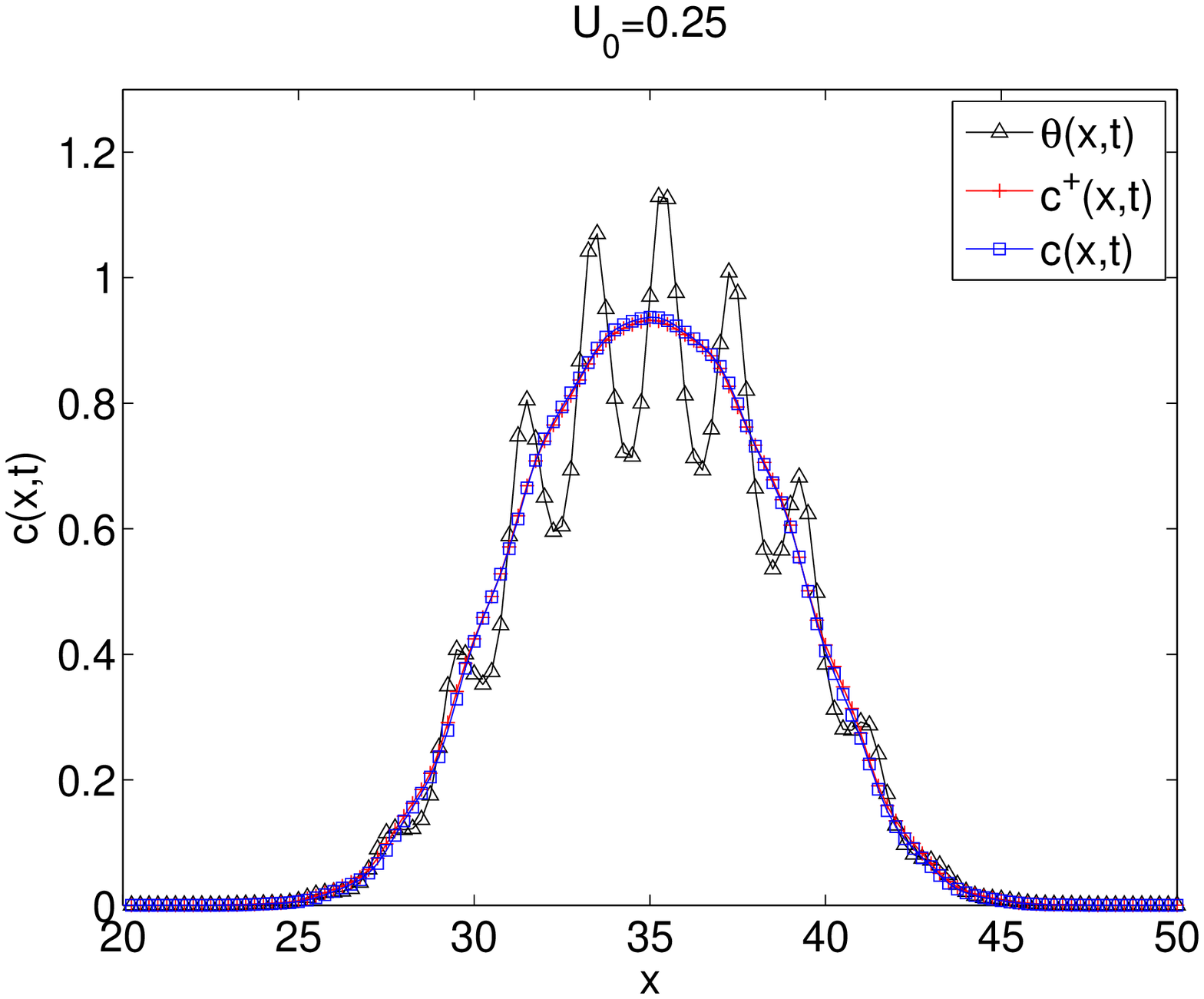}} &
      \setlength{\epsfysize}{4.2cm}
      \subfigure[]{\epsfbox{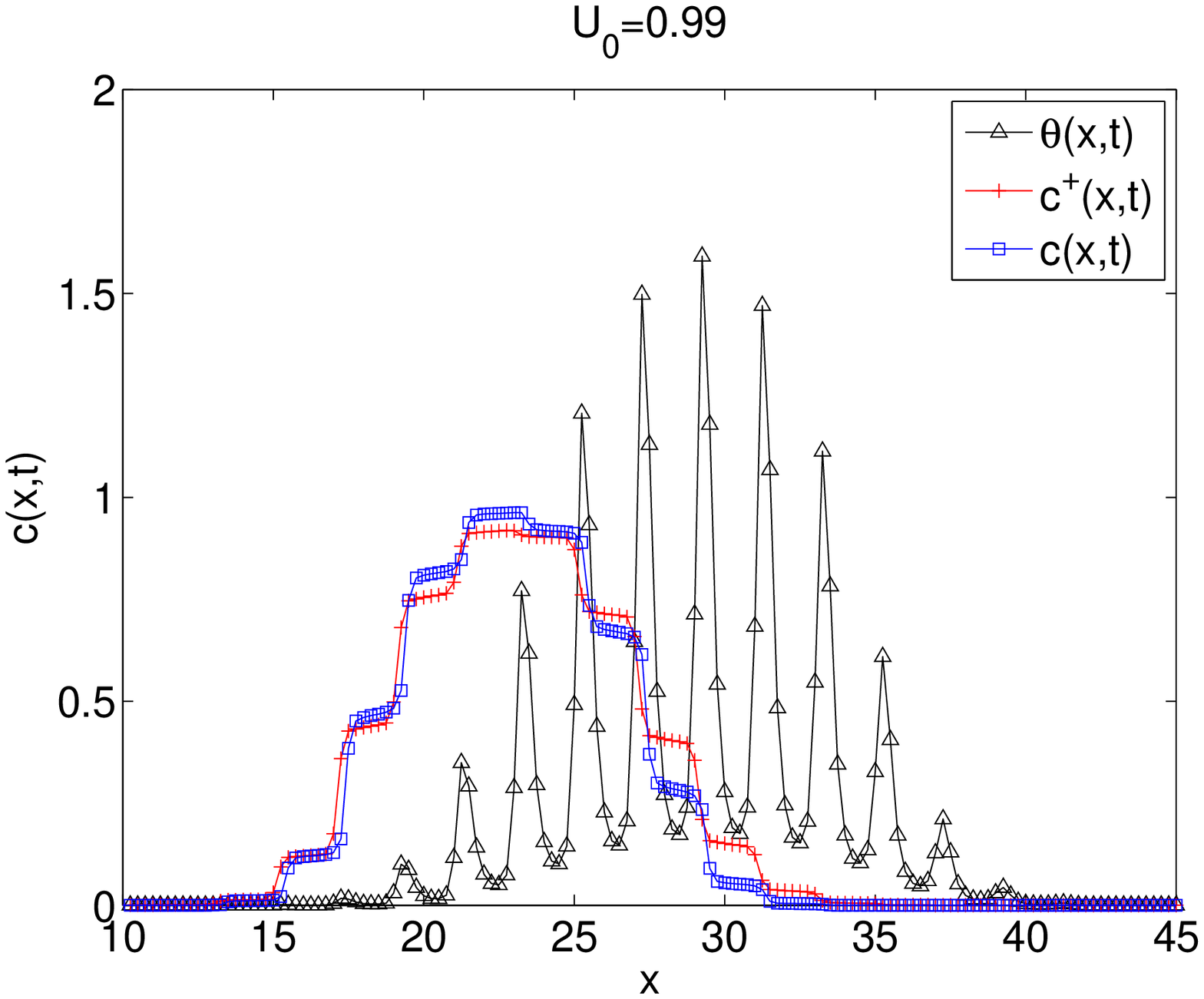}} \\[-0.2cm]
    \end{tabular}
    \caption{Comparison between results from equations (\ref{Bianco_Federicoeq10}), (\ref{Bianco_Federicoeq9}) and (\ref{Bianco_Federicoeq8})
 after a transient of $t=15$ by increasing $U_0$. The reaction rate is an FKPP type, the initial condition
  $c(x,t_0)=c^+(x,t_0)=\theta(x,t_0)$ with $x_0=20.5$ and $\sigma_0=10$, the grid step $dx=0.025$ and the time step $dt=0.0005$. For equations
(\ref{Bianco_Federicoeq10}) and (\ref{Bianco_Federicoeq9}) we set up $D=\gamma=0.1$ while for equation (\ref{Bianco_Federicoeq8}) $\rho D=\gamma=0.1$.   }
    \label{Bianco_Federicofig3}             
  \end{center}
\end{figure}

\begin{figure}[htbp]                    
  \begin{center}
    \begin{tabular}{cc} 
      \setlength{\epsfysize}{5cm}
      \subfigure[]{\epsfbox{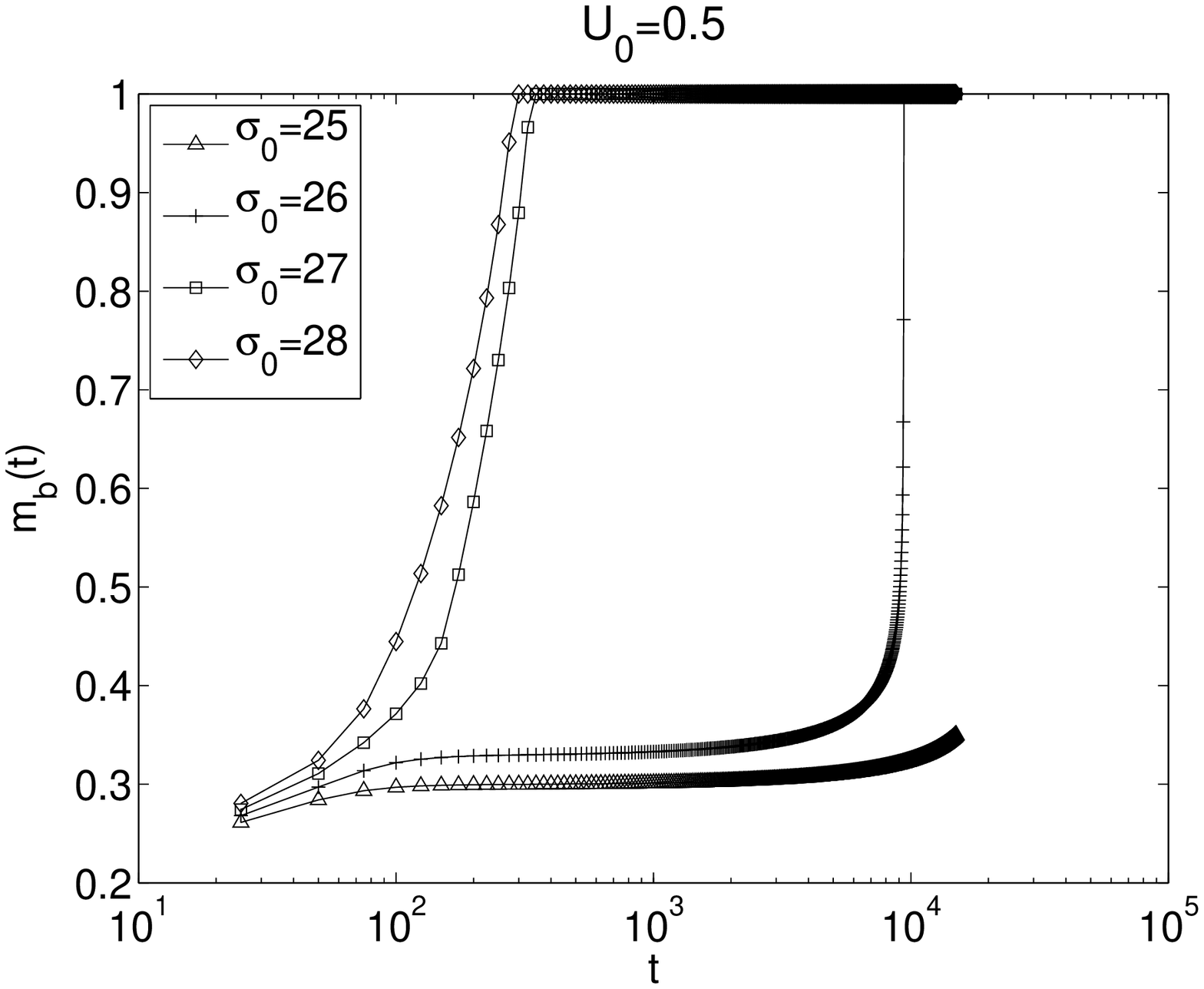}} &
      \setlength{\epsfysize}{5cm}
      \subfigure[]{\epsfbox{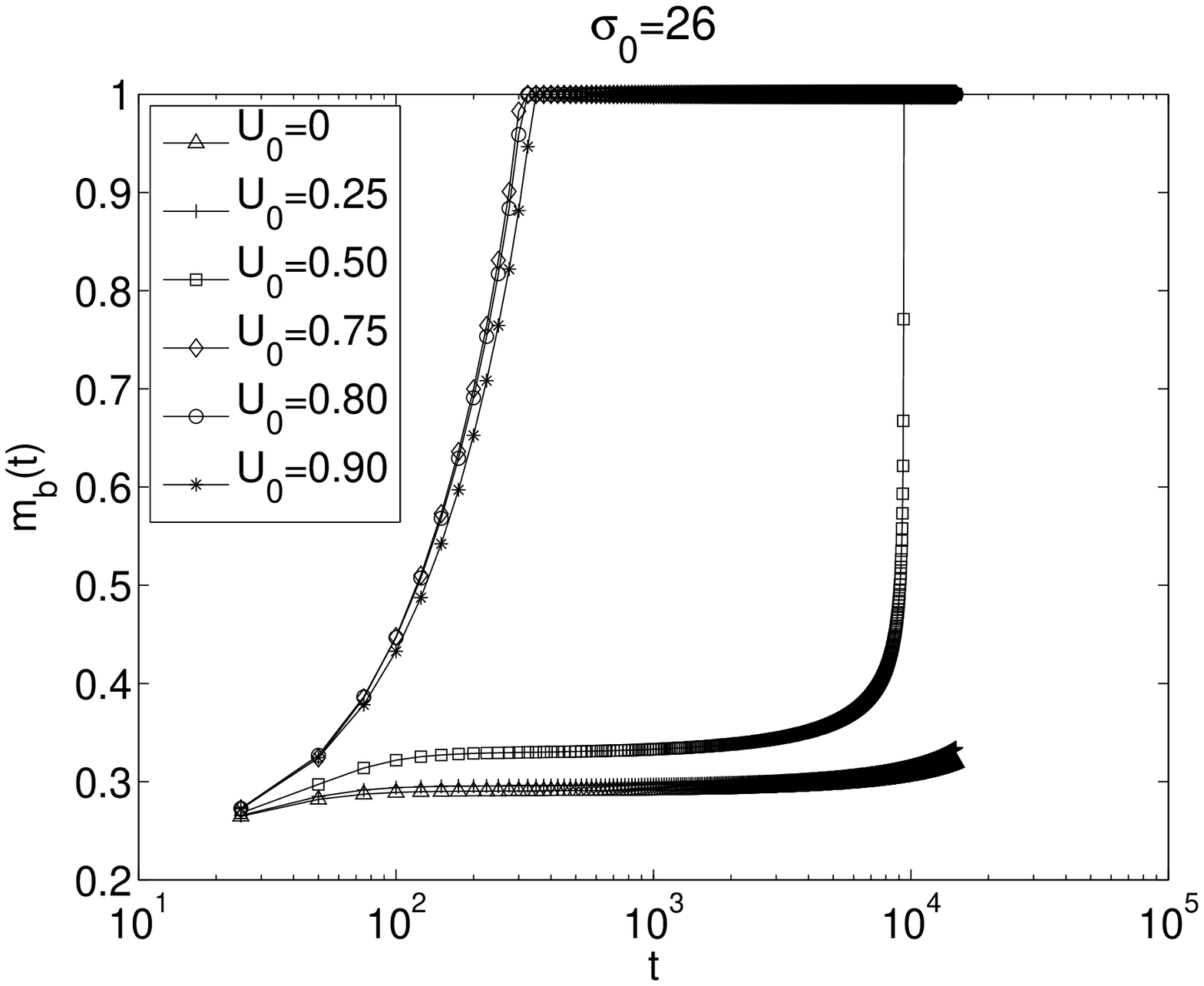}} \\[-0.2cm]
    \end{tabular}
    \caption{On subfigure (a) the mass of burnt gases ($m_b$) as function of time by increasing $m_b(t=0)$ when $U_0=0.5$.
 On subfigure (b) the mass of burnt gases ($m_b$) as function of time by increasing $U_0$ and fixed $m_b(t=0)$. 
The general mth order with $m=10$ reaction is considered. The parameter $\alpha_3$ so that $max[f_2(c)]=max[f_3(c)]$ and $\rho D=0.4$}
    \label{Bianco_Federicofig5}             
  \end{center}
\end{figure}

First of all we are interested in how the compressibility affects the time and the spatial behaviour of the fraction
of burnt gas. We will solve now equation (\ref{Bianco_Federicoeq8}) setting up an initial condition for $c(x,t)$ as:
\begin{equation}
c(x,t=0)=\exp \left[\frac{-(x-x_0)^2}{\sigma_0}\right]
\label{Bianco_Federicoeq11} 
\end{equation}
Results are shown in Fig.\ref{Bianco_Federicofig2}.
If we focus just the front interface, it is possible to identify a general behavior:
when $U_0=0$, the test case collapses to an incompressible advection-diffusion-reaction case where
$u(x,t)=1$ and $D_0=\frac{\gamma}{\beta}=\gamma$ (as $\beta=1$). Therefore, the burnt fraction 
moves with a constant velocity while it diffuses. 
The front interface is smooth end follows an exponential law.
\\By increasing $U_0$ the front interface becomes more and more uneven. When $U_0$ tends to one, it becomes discontinuous.
Moreover, the barycentric velocity of burnt fraction decreases by increasing $U_0$. 
\\These simulations were then repeated by solving equations (\ref{Bianco_Federicoeq8}) and (\ref{Bianco_Federicoeq10}) 
in the case of FKPP reaction (Fig.\ref{Bianco_Federicofig3}). We set up in both cases $D=\gamma=0.1$ and the initial condition
  $c(x,t_0)=c^+(x,t_0)=\theta(x,t_0)$.
\\The numerical method can be said to be validated since the three models are completely
 equivalent when $U_0=0$, where the three equations degenerate in the case of incompressible flow.
Moreover, increasing  $U_0$ but keeping it fairly low, there are no particular differences between the results obtained
  from equations (\ref{Bianco_Federicoeq8}) and (\ref{Bianco_Federicoeq10}).
In this range of $U_0$, the hypothesis $\rho D$  constant is not
  far from the hypothesis $D$ constant. Nevertheless results are different from those obtained through the model
 represented by equation (\ref{Bianco_Federicoeq10}).
Peaks of the concentration appear in proximity of the density peaks of the fluid. Increasing $U_0$, we increase 
the compressibility and the concentration appears more and more peaked. Here,  $ \theta $
has values higher than the unity and hence confirming that in such areas the reaction rate is negative.
For $U_0$ tending to unity, the differences between the three equations are particularly marked. The diffusion model becomes
  particularly important since it is found, as expected, a relevant difference between the results from equations
 (\ref{Bianco_Federicoeq8}) (\ref{Bianco_Federicoeq9}). 
\\We consider now the ignition reaction and we study under which conditon the reaction blows off. 
That is when the reaction stops before beeing completed.
The phenomena can be studied by following in time the mass fraction of burnt gases which is defined, in normalized form, as:
\begin{equation}
m_b(t)=\frac{\int_0^{L_x} c(x,t)\rho(x)dx}{\int_0^{L_x}\rho(x)dx}
\label{Bianco_Federicoeq12}
\end{equation}
If the reaction does not blow off, $m_b$ saturates to one. Otherwise, it stabilize to a lower value, which means that
the fractions are no more reacting.
\begin{figure}[htbp]                    
  \begin{center}
    \begin{tabular}{cc} 
      \setlength{\epsfysize}{5cm}
      \subfigure[]{\epsfbox{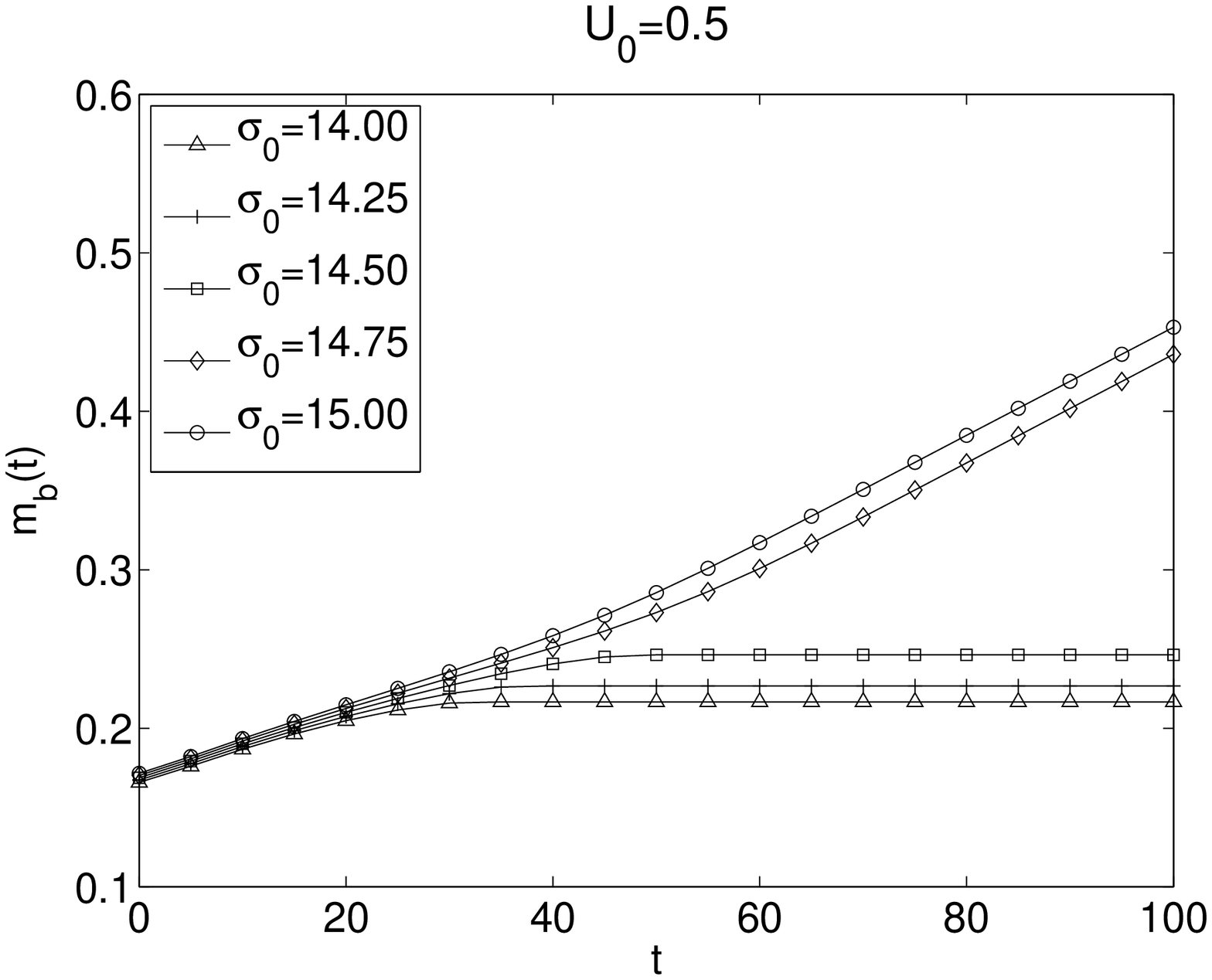}} &
      \setlength{\epsfysize}{5cm}
      \subfigure[]{\epsfbox{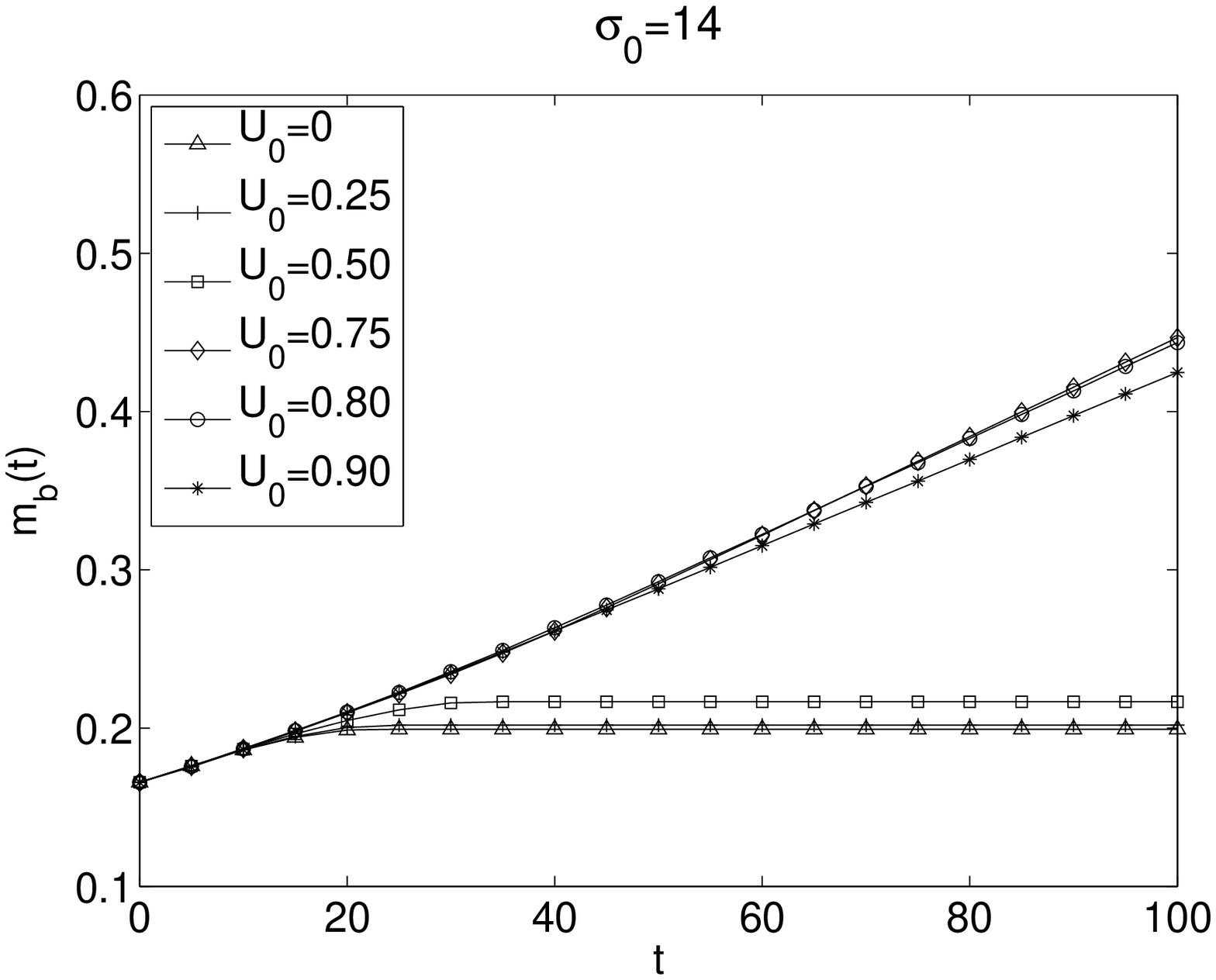}} \\[-0.2cm]
    \end{tabular}
    \caption{On figure (a) the mass of burnt gases ($m_b$) as function of time by increasing $m_b(t=0)$ when $U_0=0.5$.
 On figure (b) the mass of burnt gases ($m_b$) as function of time by increasing $U_0$ and fixed $m_b(t=0)$. 
The ignition rate of reaction with threshold $c_s=0.7$ is considered, the parameter $\alpha_2=\frac{0.1}{(1-c_s)^2}=1.111$ and the diffusion coefficient $\rho D=0.4$.}
    \label{Bianco_Federicofig4}             
  \end{center}
\end{figure}
\\ We set up an initial condition as in
(\ref{Bianco_Federicoeq11}) and we have varied $\sigma_0$, having fixed $U_0=0.5$, until the blow-off phenomena ocurr.
 Results are shown on Fig.\ref{Bianco_Federicofig4}
which confirms that, in the case of relatively high diffusion and slow reactions, if the initial burnt
 fraction is too low the reaction can quench.
Then, if we fix the initial burnt fraction, Fig.\ref{Bianco_Federicofig4} shows that by increasing the compressibility of the flow field (increasing $U_0$)
the blow-off phenomena can disappear. In paricular, there is a maxium value of $U_0\in [0.75,0.8]$, for which the reaction quenchs with more difficulty.
\\We have then studied the possibility of blow-off when the m\textsuperscript{th} order Fisher's reaction is applied. 
The approach has been the same, so that we firstly fixed $U_0=0.5$ studying a range of $\sigma_0$ for which the blow off appears and than
we fixed $\sigma_0$ studying the effects of compressibility. Results (see Fig.\ref{Bianco_Federicofig5}) shows that the rection never blows off.

\section{Conclusion}
This brief analysis has helped us to shed light on some features of the models that are typically
 used in the study of population dynamics and in combustion in the form of an advection-reaction-diffusion equation.
In particular, we showed that equation \ref{Bianco_Federicoeq6} applied to compressible flow fields leads to peaks of the concentration
$\theta$ greater then unity. This means that the rate of reaction can take negative values.
Since combustion is an irreversible process, a suitable reaction term has to be positive or zero and thus this model does not seem appropriate.
We also studied the blow off phenomenon which is typical for the ignition like reactions.
We have highlighted then the role of compressibility of the flow field.
For slow and moderate compressible flows ($U_0 < 0.8$),  the efficiency of the reaction increases. Finally, we have studied
the blow off phenomena when general mth order Fisher reaction is applied. For these kind of reactions this phenomenon does not occur.    


\vfill

\begin{thebibliography}{let1}

\bibitem{Bianco_Federico1}
{\sc Prasad Perlekar, Roberto Benzi, David R. Nelson and Federico Toschi},
Population Dynamics At High Reynolds Number
{\it Phys. Rev. Lett.}, {\bf 105} (14), 144501 (1975).

\bibitem{Bianco_Federico2}          
{\sc T. Poinsot and D. Veynante},
{\it Theoretical and numerical combustion}
Edwards (2005)

\bibitem{Bianco_Federico3}          
{\sc R. Proud'homme},
{\it Flows of reactive fluids}
Springer (2010)

\bibitem{Bianco_Federico4}          
{\sc S. Berti, D. Vergni and A. Vulpiani},
Combustion dynamics in steady compressible flows
{\it EPL}, {\bf 83}, 54003 (2008).

\bibitem{Bianco_Federico5}          
{\sc A. Kolmogorov, I. Petrovsky, N. Piscounoff},
{\it Moscow Univ. Bull. Math.}, {\bf 1} (1) (1937).




\end{thebibliography}
\end{document}